# X-RAY REFLECTIVITY MEASUREMENTS OF SPECULUM METAL MIRRORS USING SYNCHROTRON RADIATION

E.A Nazimudeen [1], T.E Girish [2]*, Sunila Abraham [1], M. H. Modi [3], M. K. Tiwari [3], C. V. Midhun [4], T.S. Shyju[5] and K. M. Varier [2]

[1] Department of Physics, Christian College, Chengannur, Kerala, India – 689122
[2] Department of Physics, University College, Thiruvananthapuram, Kerala, India - 695034
[3] Synchrotron Utilization Section, RRCAT, Indore, Madhya Pradesh, India - 452 013
[4] Department of Physics, University of Calicut, Malappuram, Kerala, India – 673635
[5] Centre for Nanoscience & Nanotechnology, Sathyabama Institute of Science and Technology (Deemed to be University),Chennai,India - 600119

*Author for correspondence: tegirish5@yahoo.co.in

**Abstract**

Recent advances in grazing incidence X-ray optics using synchrotron radiation sources have stimulated the need for basic research in high quality mirror materials for novel applications. In this paper we communicate the results of the first measurements of glazing angle X-ray reflectivity (XRR) of speculum metal mirrors using synchrotron radiation sources. Our results agree with similar measurements made in polished speculum gratings by Arthur Compton and his collaborators during 1923 using ordinary X-ray tubes. Our experimental investigations are based on synchrotron radiation research facilities (Indus 2) maintained by Govt of India in Indore . Variations in the XRR with grazing angles of incidence and X-ray energy for cast, thin film and electron irradiated samples of speculum metal will be discussed in this paper.

## 1. Introduction

Speculum metal is a binary alloy of Copper and tin in the proportion of about 2:1, whose antiquity dates backs to Bronze Age [1]. It is used as a optical material from the time of Isaac Newton in 17th century and continuing up to early decades of 20th century in mirrors, gratings, large reflecting telescopes and other optical precession instruments [2]. Speculum metal finds excellent applications in diverse fields of science and technology apart from its archaeometallurgical importance. It has several engineering applications in various fields such as slit material in transmission electron microscopy, alternative to nickel undercoating for gold or chromium electroplating, anode material for lithium- ion batteries, etc. [3-4]. Speculum metal was part of modern scientific observations and discoveries covering different areas including experimental astronomy, optical and nuclear instrumentation, spectroscopic and other studies in modern physics. Materials science properties [5] and hard X-ray reflectivity of speculum metal mirrors are recently reported [6].

Recent advances in grazing incidence reflecting optics has transformed the field of observational X-ray astronomy into a major scientific discipline in astrophysics and cosmology. Space astronomy has revolutionized our view of the invisible universe in wave bands that shed light upon dark energy and dark matter. This development has been triggered by the availability of high-brilliance synchrotron radiation sources that stimulated the stimulated the need for basic research in high quality optically active mirror materials for novel applications [8-11]. Thus, we have studied the grazing incidence X-ray reflectivity of speculum metal by using microprobe X-Ray Fluorescence beamline of synchrotron radiation source- Indus 2. Variations in the reflectivity of cast, thin film and electron irradiated samples of speculum metal are also discussed as functions of grazing angles of incidence at different photon energies of 8 keV, 17.5 keV and 20 keV.

3. **Materials and methods**

Experimental investigations are carried out by using a highly reflecting speculum metal mirror, cast and polished using traditional techniques from Aranmula, a village in Kerala, Southern India. There exist literary articles that describe its metallurgical and technical aspects of casting and polishing methods [12-14]. The purchased mirror (Aditi Handicrafts Centre, Aranmula) with an aperture of 0.05m, separated from the brass base is cleaned and sectioned into number of fragments of different dimensions using diamond cutter, which are used for experimental characterization. Speculum thin films were prepared on glass substrate by the method of thermal evaporation. The speculum powder containing approximately 65% by weight of Cu and 32% by weight of Sn was loaded in a molybdenum crucible placed in vacuum and evaporation was carried out at room temperature by applying a current of about 120A with a coating pressure of about $2 \times 10^{-5}$ mbar in a coating time of around 5min. Nuclear irradiations at the polished surface of cast speculum metal samples were carried out by using Varian clinac-iX medical accelerator having Nominal potential of 20MeV with a field size of 10x10 cm2 and irradiation distance 100 cm from bremsstrahlung target. Electron beam irradiations with energies 6 MeV and 9 MeV were applied at the polished surface of speculum samples with a dose rate of 400 cGy/min and current of 20 µA. X-ray reflection measurements were carried out in θ- 2θ geometry using Bl-16 reflectometer station at Indus 2 synchrotron facility in the ambient air environment. Collimated X-rays of 8 keV, 17.5 keV and 20 keV, monocromatized1 using a Si (1 1 1) double crystal monocromater were used for the measurement. Fig. 1 illustrates a schematic optical layout of the BL-16 beamline, while Fig. 2 depicts a photograph of the installed x-ray reflectometer station at BL-16 beamline [15, 16].

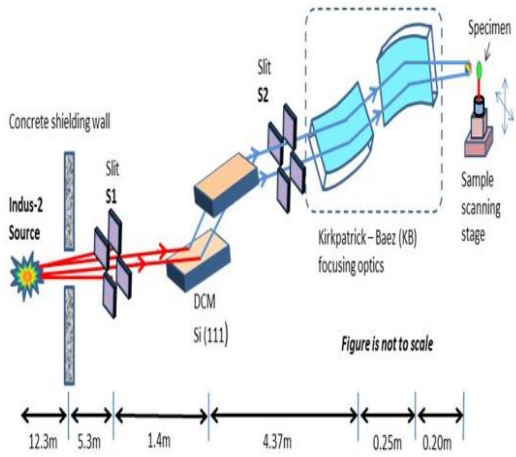
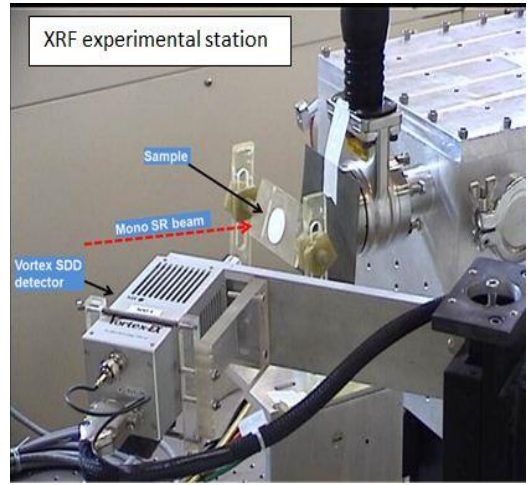

Fig. 1  A schematic optical layout of the BL-16 beamline showing different components of the beamline [RRCAT, Indore].

Fig. 2.  A photograph of the x-ray reflectometer station at BL-16 beamline [RRCAT, Indore

## 4. Results and discussion

Experiments related to X- ray optics carried out by A. H. Compton and his collaborators during 1923 – 1927, led to the Nobel Prize winning discovery of Compton Effect [17-19]. Based on the critical analyses of these experimental investigation results, we could infer a unique X- ray reflection property of speculum metal that provided us the evidence for its reflectivity in the hard X-ray region. Reflection on speculum metal using Cu $K_\alpha$ and Mo $K_\alpha$ X-rays with energy 8 keV and 17 keV are shown in Fig. 3. In connection with this, the X-ray reflectivity of cast, thin film and electron irradiated samples of speculum metal were studied experimentally as a function of grazing angle of incidence ranging from 0 - 2 ° with a step size of 0.01◦ at three different photon energies of 8 keV, 17.5 keV and 20 keV using the BL-16 beamline of the National synchrotron radiation light source Indus 2 at RRCAT, Indore.

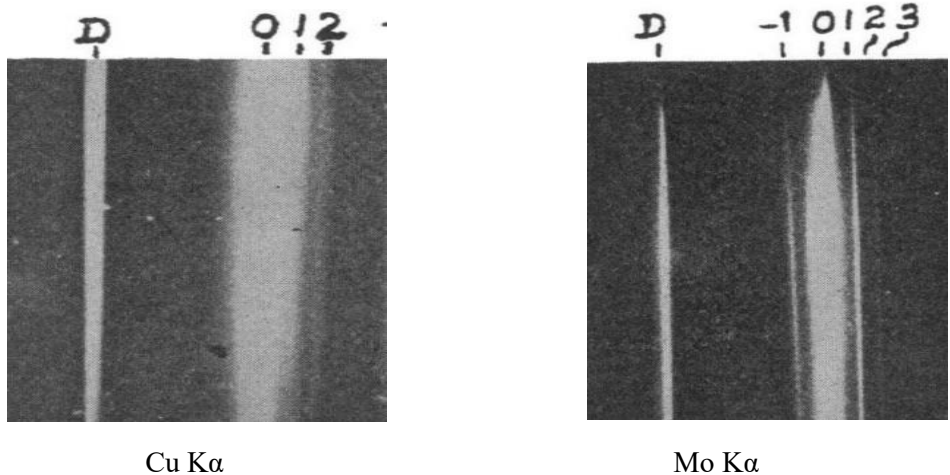

Cu Kα                                    Mo Kα

**Fig. 3. Reflection on polished speculum metal gratings using Cu Kα and Mo Kα X- rays as reported by Compton in 1923 [18]. Here D – Image of direct beam & O - reflected beam**.

The dependence of x-ray reflectivity of cast, thin film and electron irradiated samples on incidence angle for three different x-ray energies are illustrated in Figs.5- 8. Total external reflection critical angle of speculum metal for 8 keV, 17.5 keV and 20 keV X-rays are about 0.4, 0.18 and 0.16 respectively. It is also observed from all plot that the reflectivity is high at grazing angle incidence up to the critical angle and beyond that the reflectivity falls rapidly down to almost zero. The reduction in reflectivity is a strong indication of the direct function of surface roughness. Electron irradiated and thin film samples show high value reflectivity (90%) in comparison with cast speculum metal (80%). At 8 keV X-ray, all the three samples shows high reflectivity (above 80 %) at grazing angle ranging from 0.2- 0.4 °, but with other two x-ray energies, high reflectivity is observed within small range of incidence angle. This high reflectivity of speculum metal at 8 keV and 17.5 keV obtained from synchrotron based X-ray reflection analysis confirms the results of Compton and Deon experiments (1923) on the X-ray reflectivity of speculum metal using Cu Kα and Mo Kα, which is about 80- 90 %.

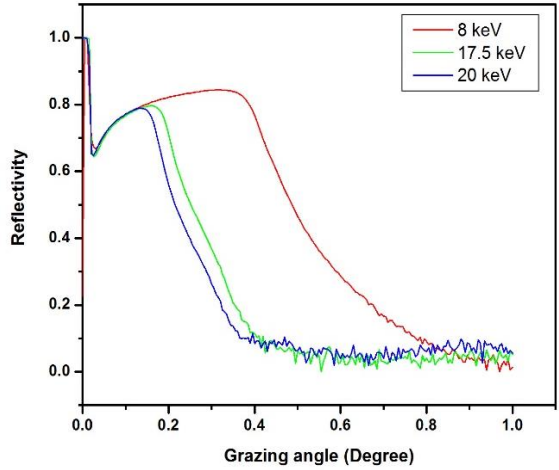

**Fig. 4** Experimental X-ray reflectivity of cast speculum metal as functions of grazing angle of incidence at photon energies of 8 keV, 17.5 keV and 20 keV

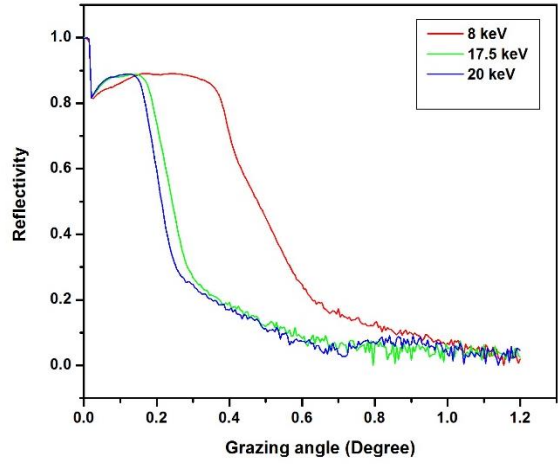

**Fig. 4** Experimental X-ray reflectivity of thin film speculum metal as functions of grazing angle of incidence at photon energies of 8 keV, 17.5 keV and 20 keV

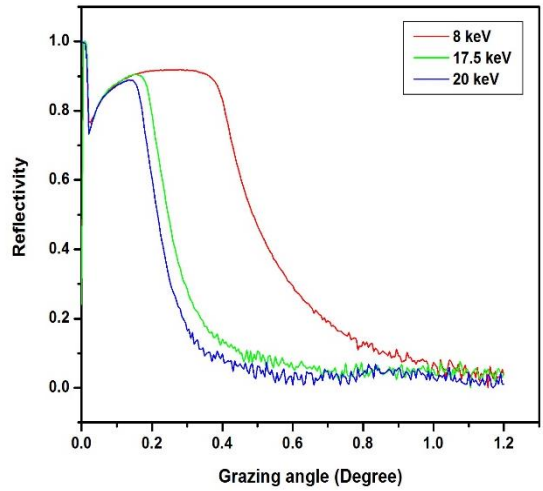

**Fig. 6** Experimental X-ray reflectivity of 6 MeV electron irradiated cast speculum metal as functions of grazing angle of incidence at photon energies of 8 keV, 17.5 keV and 20 keV

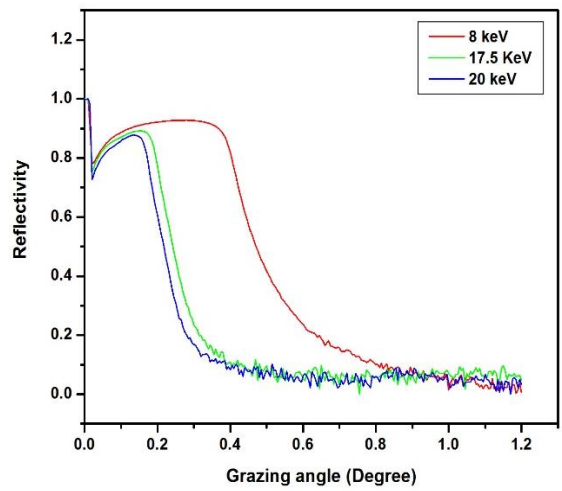

**Fig. 7** Experimental X-ray reflectivity of 9 MeV electron irradiated cast speculum metal as functions of grazing angle of incidence at photon energies of 8 keV, 17.5 keV and 20 keV

## 5. Conclusions

X-ray reflectivity of cast, thin film and electron irradiated samples of speculum metal are studied as functions of grazing angle at different photon energies of 8 keV, 17.5 keV and 20 keV using synchrotron based microprobe XRR technique. Thin film and electron irradiated speculum metal samples show high X-ray reflectivity (about 90%) compared to cast material (about 80 %). Synchrotron based microprobe XRR analysis confirmed the results of Compton and Doan experiments (1923) on the Grazing angle X-ray reflectivity of speculum metal, which is about 80- 90 %. The present results may open up promising possibilities for this alloy as mirrors, gratings, filters in various fields of nuclear, medical and astrophysics especially in X-ray astronomy, X-ray microscopy, X-ray imaging, synchrotron beamlines and other areas of high energy physics.

**Acknowledgements**

Authors are extremely grateful to the Director and Dr.Tapas Ganguly for granting permission to use Indus-2: beam -16 research facilities for measurements of grazing angle X-ray reflectivity studies and for our stay in RRCAT,Indore. We also thank Dr. Ajith Sing, Mr. Ajay and other faculty members and scientists in this centre for their kind help, constructive comments and valuable suggestions during our experimental investigations.

**References**

[1]  William F Collins, The mirror- black and quicksilver patinas of certain Chinese bronzes. *J. R. Anthropol. Inst. G*. 64 (1934) 69- 79. http://www.jstor.org/stable/284394

[2]  A. A. Mills, R. Hall, The production of a plane surface: As illustrated by specula from some early Newtonian telescopes, *Notes Rec. Roy. Soc. London*. 37 (1983) 147- 166. http:// doi: 10. 1098 /rsnr.1983 .0008

[3]  A.V. Crewe, M. Isaacson, D. Johnson, *Rev. of Sci. Instrum*, 1971, 42, 411- 420.

[4]  Akira Ito, KuaikiMurase, Takashi Ichii, Hiroyuki Sukimura, *ECS transactions,* 2009, 16, 677- 679.

[5]  E.A. Nazimudeen, T. E. Girih, Sunila Abraham, On the nature of thin films of speculum metal, *Met. comms* 16 (2018) 314- 319.https://doi.org/10.1016/j.mtcomm.2018.07.005

[6] Nazimudeen E.A, T.E. Girish, Sunila Abraham, Early Compton Effect experiments revisited: Evidence for outstanding hard X-ray reflectivity of speculum metal (2019) *arXiv:2001.02053v1 [cond-mat.mtrl-sci]*.

[7] Jianguo Ren, Xiangming He, Liwang, Weihua Pu, Changyin Jiang, Chunrong Was, *ElectrochimicaActa*, 2007 52, 2447- 2452.

[8] Paul Gorenstein, Grazing incidence telescopes for x-ray astronomy, Opt. Eng. 51(2012) 011010: 1-14.

[9] A. Vikhlinin et al., Chandra cluster cosmology project iii: cosmological parameter constraints, Astrophys. J. 692 (2009) 1060.

[10] S. W. Allen et al., Improved constraints on dark energy from Chandra x-ray observations of the largest relaxed galaxy clusters, Mon. Not. R. Astron. Soc. **383**, 3 (2008), 879–896.

[11] A. M. Soderberg, E. Berger, K. L. Page et al., An extremely luminous x-ray outburst at the birth of a supernova, Nature. 453, 7194 (2008) 469–474.

[12] S. Srinivasan, I. Glover, Skilled mirror craft of intermetallic delta high-tin bronze ($Cu_{31}Sn_8$, 32.6 % tin) from Aranmula, Kerala, Current Sc.93 (2007) 35-40.

[13] J.A Sekhar et al., Ancient metal mirror alloy revisited: Quasicrystalline nanoparticles observed, JOM. 67(2015) 2976- 2983.

[14] T. Nagee et al., Thermographical analysis of continuing tradition of mirror casting in Kerala, ISIJ International.54 (2014) 1172 – 1176.

[15] Gangadhar Das et al., Simultaneous measurements of X-ray reflectivity and grazing incidence fluorescence at BL-16 beamline of Indus-2, Review of Scientific Instruments 86, 055102 (2015); doi: 10.1063/1.4919557

[16]  M. K. Tiwari, Gangadhar Das, An Interactive Graphical Use Interface (GUI) for the CATGIXRF Program- for Microstructural Evaluation of Thin Film and Impurity Doped Surfaces, X-Ray Spectrom. 2016, 45, 212–219, DOI 10.1002/xrs.2692

[17]  A. H. Compton, R.L. Doan, X- ray spectra from a ruled reflection grating. Proc. N.A.S. 11 (1925) 596-601.

[18]  A. H. Compton, The total reflection of X- rays, Phil. Mag. S. 6. **45** (1923) 1121- 1131

[19]  A. H. Compton, S. K. Allison, X-rays in Theory and Experiment, second ed., D. Van Nostrand Company, Inc., New York, 1935.